\documentclass[conference]{IEEEtran}

\usepackage{cite}

\usepackage[pdftex]{graphicx}
\usepackage{wrapfig}

\ifCLASSOPTIONcompsoc 
	\usepackage[caption=false,font=normalsize,labelfont=sf,textfont=sf]{subfig}
\else
	\usepackage[caption=false,font=footnotesize]{subfig}
\fi

\usepackage{algorithm}
\usepackage{algpseudocode}

\usepackage{booktabs}
\usepackage[table,xcdraw]{xcolor}

\usepackage{alltt}
\usepackage[flushleft]{threeparttable}
\usepackage{multirow}
\usepackage{hhline}

\begin{document}
%
\title{Accelerating DNA Sequence Analysis using Intel\textsuperscript{\textregistered} \\Xeon Phi\texttrademark}

\author{\IEEEauthorblockN{Suejb Memeti and Sabri Pllana}
\IEEEauthorblockA{Linnaeus University, Department of Computer Science,\\
	351 95 V\"{a}xj\"{o}, Sweden\\
  Email: \{suejb.memeti, sabri.pllana\}@lnu.se
  }
}

\IEEEspecialpapernotice{\footnotesize(PBio at ISPA-2015, \copyright IEEE)}

\maketitle

\begin{abstract}
Genetic information is increasing exponentially, doubling every 18 months. Analyzing this information within a reasonable amount of time requires parallel computing resources. While considerable research has addressed DNA analysis using GPUs, so far not much attention has been paid to the Intel Xeon Phi coprocessor. In this paper we present an algorithm for large-scale DNA analysis that exploits thread-level and the SIMD parallelism of the Intel Xeon Phi. We evaluate our approach for various numbers of cores and thread allocation affinities in the context of real-world DNA sequences of mouse, cat, dog, chicken, human and turkey. The experimental results on Intel Xeon Phi show speed-ups of up to $10\times$ compared to a sequential implementation running on an Intel Xeon processor E5.
\end{abstract}

\IEEEpeerreviewmaketitle

\section{Introduction} \label{introduction}

There is a growing interest in molecular biology community to understand the information that is encoded within the Deoxyribonucleic Acid (DNA) sequences of each organism \cite{Editorial}. A DNA sequence contains specific genetic instructions that make the living organisms function in a proper way. The four basic building blocks (also known as \emph{nucleotide bases}) of a DNA sequence are: \emph{Adenine} (A), \emph{Cytosine} (C), \emph{Guanine} (G) and \emph{Thymine} (T). 

Discovery of differences and similarities of organisms and exploration of the evolutionary relationship between them, often require comparisons of the corresponding DNA sequences. Examples include: checking whether one sequence is a sub-sequence of another, or finding a sub-sequence that appears in the same order in both DNA sequences \cite{Bentley01102000}. The process of searching for certain sub-sequences of length \emph{k}, so called \emph{k-mers}, is performed with pattern matching algorithms. 

According to Benson et al. \cite{benson2013genbank} the number of DNA sequences and nucleotide bases in these sequences is doubling every 18 months. Real-world DNA sequences comprise several Gigabytes and the process of extracting the important information demands the adequate use of parallel computing resources to be completed within a reasonable time. A quick DNA analysis may have a decisive role in many applications including: preventing the evolution of different viruses and bacterias during an early phase \cite{collins2003vision}; early diagnosis of genetic predispositions to certain diseases (such as, cancer, cardiovascular diseases,..) \cite{mellmann2011prospective}; and DNA forensics (such as, parentage testing, or criminal investigation) \cite{luftig2000dna}.

Related research has addressed extensively pattern matching algorithms for GPUs. Lin et al. \cite{Lin_PFA-C_Algo} proposed the Parallel Failure-less Aho-Corasick algorithm for pattern matching on GPUs. Kouzinopoulos and Margaritis \cite{kouzinopoulos2009string} show speedup of up to 24$\times$ for small input text and pattern sizes for different algorithms on GPU. Bellekens et al. \cite{bellekens2013investigation} presented a parallel implementation of the Knuth-Morris-Pratt algorithm using the Nvidia GPU hardware. Tumeo and Villa \cite{Tumeo_DNA_GPU} proposed an implementation of the Aho-Corasick algorithm for DNA analysis applications on clusters with GPUs. In comparison to GPUs, besides the ability to provide high performance, the Xeon Phi deserves our attention because of programmability \cite{DokulilBBPSB13,PllanaBMNX08} and portability \cite{KesslerDTNRDBTP12}. However, so far not much research was focused on DNA analysis using pattern matching algorithms designed specifically for the Xeon Phi.

In this paper, we present a parallel algorithm for DNA analysis that is designed to exploit the thread-level and the SIMD parallelism available in the Intel Xeon Phi coprocessor. Our pattern matching algorithm is based on finite automata. For thread-level parallelism we use a domain decomposition approach that splits the DNA sequence into chunks evenly among the available threads. To process the patterns occurring in the cross border of sequence chunks, our algorithm uses $m-1$ overlapping characters, where \emph{m} is the pattern length. With respect to the SIMD-parallelism our algorithm implementation uses the potential of the 512-bit vector registers of the Intel Xeon Phi architecture for transition function of finite automata. We evaluate our approach experimentally with real-world DNA sequences of different living species. For the human DNA sequence a speedup of up to $10\times$ is achieved compared to the sequential version running on Intel Xeon processor E5. Major contributions of this paper include,
\begin{itemize}
	\item an algorithm for large-scale DNA analysis that is designed for Intel Xeon Phi;
	\item an experimental evaluation of our DNA analysis algorithm for real-world DNA sequences of mouse (2.7GB), cat (2.4GB), dog (2.4GB), chicken (1GB), human (3.2GB) and turkey (0.2GB);
	\item a discussion of the state-of-the-art in pattern matching and DNA analysis using many-core architectures (GPUs, Intel Xeon Phi).
\end{itemize}


The rest of the paper is organized as follows. Section \ref{background} provides background information with respect to pattern matching and introduces the Intel Xeon Phi architecture. Our parallel algorithm for DNA analysis using the Intel Xeon Phi coprocessor is described in Section \ref{our-algorithm}. Section \ref{exp_evaluation} presents the experimental setup and discusses the experimental results. The work described in this paper is compared and contrasted to the state-of-the-art related work in Section \ref{related_work}. Section \ref{summary_future_work} provides a summary of this paper. 


\section{Background} 
\label{background}

In this section we first provide background information with respect to the pattern matching with finite automata. Thereafter, we present the major features and the architecture of the Intel Xeon Phi coprocessor.

\subsection{Pattern Matching with Finite Automata (FA)}

Finding occurrences of a pattern in a text is a frequent need of many text-editing programs (e.g. find-replace functions), Internet Search Engines (e.g. finding web-pages that are relevant to the provided query), or lexical analyzers (e.g. determining the locations of a pattern within a sequence of tokens). In the context of computational biology, pattern matching algorithms are used for analyzing and processing genetic information by searching for particular patterns in DNA sequences. Formally, in DNA analysis the string matching problem can be expressed as follows: the input text (DNA sequence) is an array $T[1..n]$ where $n$ is the length of the DNA sequence, and pattern $P[1..m]$ where the length of the pattern $m \leq n$. The finite alphabet $\sum$ defines the possible characters of $T$ and $P$, in this case $\sum = \{A,C,G,T\}$, where each letter corresponds to one of the four nucleotide bases \cite{hopcroftullman}.

A Finite Automata (FA) is a simple machine for processing information, which scans the input text $T$ in order to find the occurrences of the pattern $P$. FA is an efficient technique for pattern matching, because it examines each character from $T$ exactly once. Formally, FA is a quintuple of $(Q, \sum, \delta, q_0, F)$, where $Q$ is a finite set of states, $\sum$ is a finite input alphabet, $\delta$ (\emph{transition function}) is the function $Q \times \sum \rightarrow Q$, $q_0$ is the start state and $F$ is a distinguished set of accepting states \cite{hopcroftullman}.

A well known algorithm for detecting any exact occurrences (including the overlapping ones) of multiple patterns is the Aho-Corasick (AC) algorithm \cite{Aho-Corasick}. Because of the capability to deliver input-independent performance, we use the AC algorithm as a basis of our algorithm for counting and extracting patterns from a DNA sequence. The AC algorithm builds an automaton by creating states and transitions corresponding to the states. The automaton is able to match multiple and overlapping occurrences, by adding a failure transition when there is no regular transition leaving from the current state. The failure transitions, known as $\epsilon$-transitions, do not consume any input, which make the automaton non-deterministic.


\subsection{Intel\textsuperscript{\textregistered} Xeon Phi\texttrademark ~Architecture}

\begin{figure}[!ht]
	\centering
	\includegraphics[width= 0.9 \linewidth]{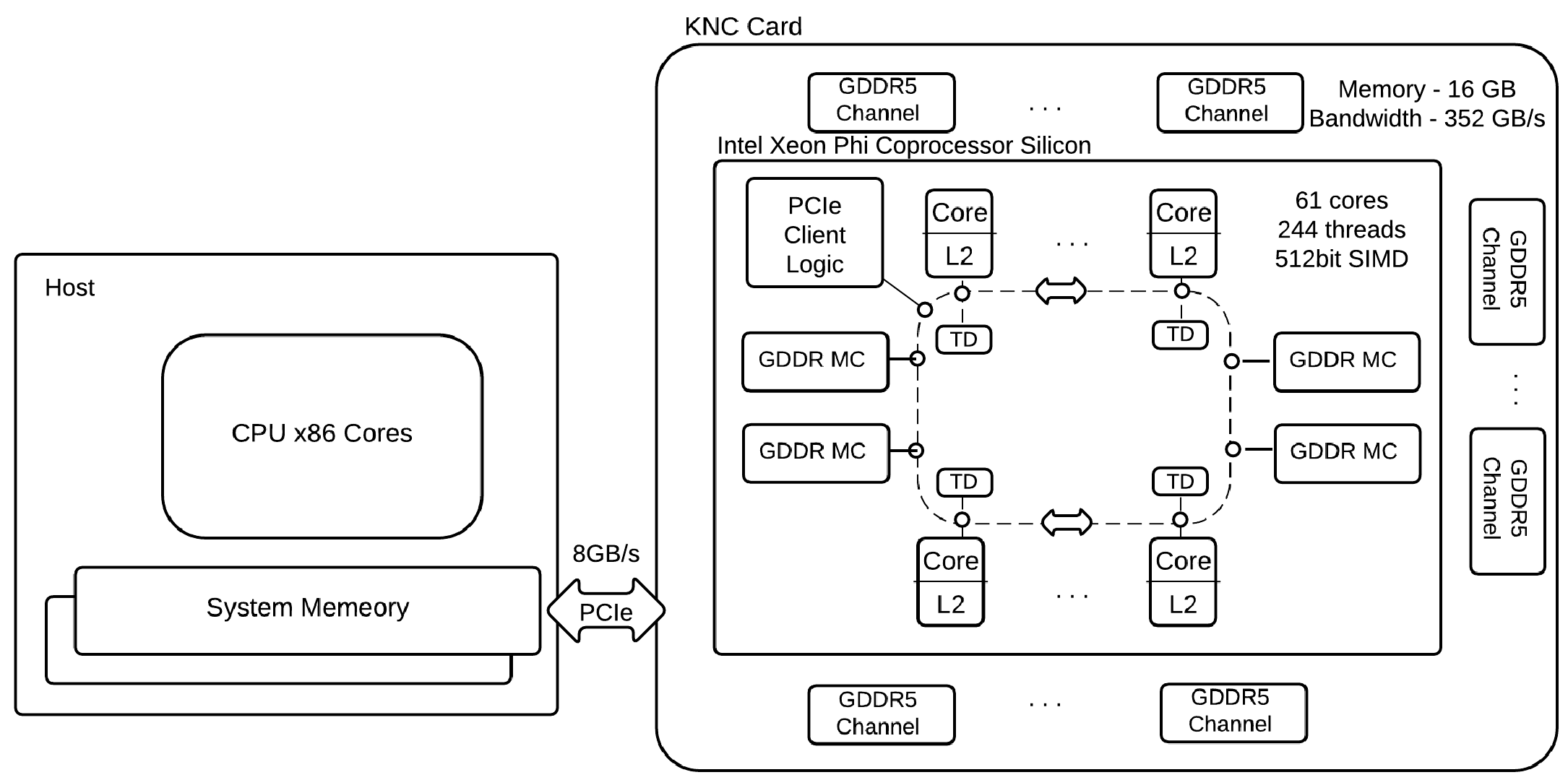}
	\caption{The Xeon Phi Architecture}
	\label{fig:xphi-platform}
	\vspace{-10pt}
\end{figure}

The Intel Xeon Phi (codenamed Knights Corner) is a many-core shared-memory coprocessor, which runs a lightweight Linux Operating System. In this paper we use the Intel Xeon Phi coprocessor 7120P. Figure \ref{fig:xphi-platform} depicts the architecture of our platform, where in the left-hand side is the host comprising one or more Intel x86 CPUs, whereas the right-hand side depicts the Intel Xeon Phi architecture. The Xeon Phi comprises 61 x86 cores, each running at 1.2 GHz base frequency with the max turbo frequency 1.3 GHz \cite{chrysos2014intel}. Each core has four hardware threads, in total there are 244 hardware threads per coprocessor, capable of delivering performance up to two TeraFLOP/s at single precision or one TeraFLOP/s at double precision. Each core has a private L2 cache of 512KB that is kept fully coherent by a global-distributed tag directory (TD). The L2 caches are connected through a bidirectional ring bus interconnect, which forms a unified shared L2 cache of 30,5MB. In addition to the cores, there are 16 memory channels, which theoretically deliver up to 352 GB/s memory bandwidth. The memory controllers (GDDR MC) and the PCIe Client Logic provide a direct interface to the GDDR5 memory and the PCIe bus, respectively. The host communicates with the coprocessor through the PCIe bus which is limited to 8GB/s transfer bandwidth. The PCIe bus is a bottleneck for the offload programming model, where data has to be transferred from the host to the coprocessor and vice versa. In order to achieve high offload computational performance, it is recommended that the data is transfered to the coprocessor and kept there (reused) to avoid memory bandwidth bottlenecks while moving the data back and forth.

An important aspect of the coprocessor is its vector processing unit, which feature Intel Advanced Vector Extensions (AVX) 512-bit SIMD instruction set. Thus it can execute 16 single-precision (16 wide $\times$ 32 bit) or 8 double-precision (8 wide $\times$ 64 bit) operations per cycle. Exploiting the vector units in an efficient way is one of the key aspects in achieving high performance on Intel Xeon Phi Coprocessor \cite{TianSPGKMCP13}.


\section{Design and Implementation of an Algorithm for DNA Analysis using Intel Xeon Phi}
\label{our-algorithm}

The key features of our algorithm (Section~\ref{sec:alg1}) and implementation for parallel DNA analysis on Intel Xeon Phi are: (1) decomposition of the input DNA sequence across the available threads, (2) exploiting the SIMD parallelism, and (3) reducing the memory references using a suitable representation for the State Transition Table (Section~\ref{sec:impl}).

\subsection {Parallel DNA Analysis Algorithm}
\label{sec:alg1}

\begin{algorithm}[t]
	\caption{Parallel DNA Analysis}\label{alg:pac}
	\begin{algorithmic}[1]
		\Require {$STT$, final state $f$, input $T$, number of threads $p$, vector length $v$, pattern length $m$}
		\Ensure Count of pattern matches and their location
		\Procedure{ac}{$dfa, f, T, p, v, m$}
		\State {$n = T.length$}
		\State {$chunkLength = n / p$}
		\State {$vChunkLength = (chunkLength + m-1)/v + m-1$} \label{alg:pac:split}
		\For{$i = 1$ \textbf{to} $p$}
		\State {$q[v]$}\Comment {store the current state for each SIMD chunk}
		\State {$chunkStart = i * chunkLength$}
		\For{$j = 1$ \textbf{to} $vChunkLength$}
		\State {$vStart = j * chunkStart$} \label{alg:pac:init_v}
		\For{$k = 1$ \textbf{to} $v$}\label{alg:pac:simd_loop} \Comment{this loop is vectorized}
		\State {$c = v*k+vStart$}
		\State {$q[k] = \delta(q[k], T[c])$} \Comment{load the next node from dfa, by following the transition from the current node $q[k]$ labeled by the symbol $T[c]$} \label{alg:pac:load_next_node}
		\If {$q[k] \geq f$} \Comment{check if the transition to next node is final}\label{alg:pac:check_final}
		\State {\textbf{print} matching pattern at position $c$}
		\EndIf
		\EndFor
		\EndFor
		\EndFor
		\EndProcedure
	\end{algorithmic}
\end{algorithm}

\begin{figure}[ht]
	\label{fig:paradna}
	\centering
	\subfloat[]{
		\includegraphics[width=0.4\linewidth]{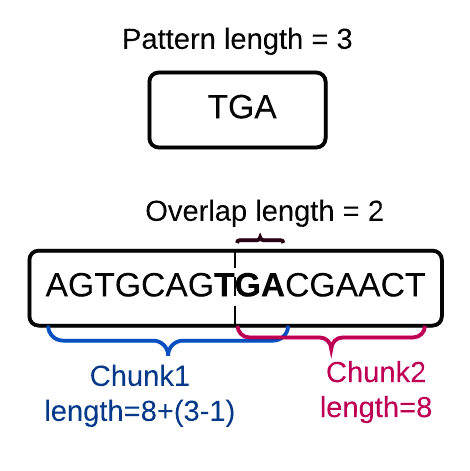}
		\label{fig:split}
	}
	\hfil
	\subfloat[]{
		\includegraphics[width=0.3\linewidth]{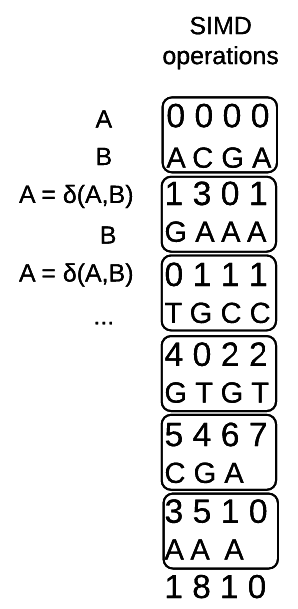}
		\label{fig:simd}
	}
	\caption {Thread-level and SIMD parallelism; (a) splitting the DNA sequence into chunks; (b) vectorization of the transition function.}
	\label{fig_sim}
\end{figure}

Fig. \ref{fig:split} illustrates our parallelization strategy that is based on domain decomposition, which means the input DNA sequence is evenly split into chunks among the available threads (Alg.~\ref{alg:pac}, lines 5 -- 18). While splitting the input, there is a risk of not being able to match the occurrences of patterns that cross the chunks boundaries. Other researchers have addressed this issue by using speculation based on most visited states \cite{LuchaupSEJ11}, by using Suffix-Arrays \cite{Chacon} or using an intersection of successor states and predecessor states of the FA \cite{memeti2014}. In this paper we find these occurrences by overlapping the input chunks by $m-1$ characters (Alg. \ref{alg:pac} Line \ref{alg:pac:split}), where $m$ is the pattern length. This method is applicable to multiple patterns with equal length, otherwise it can happen that two threads match the same pattern with a length shorter than $m-1$.

Our algorithm exploits vector units of Intel Xeon Phi, by splitting the chunks further into $v$ parts where $v$ represents the vector length (Alg.~\ref{alg:pac}, lines 10 -- 16). The operations (such as, determining the next state (Alg. \ref{alg:pac}, Line \ref{alg:pac:load_next_node}), or checking if the next state is a final one (Alg. \ref{alg:pac}, line \ref{alg:pac:check_final})) are performed on multiple data points simultaneously. 

\begin{figure}[h]
	\centering
	\includegraphics[width=0.9\linewidth]{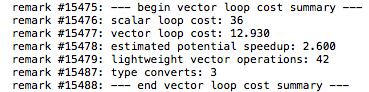}
	\caption{Vectorization report}
	\label{fig:vec-report}
	\vspace{-10pt}
\end{figure}

Fig. \ref{fig:simd} illustrates the SIMD operations assuming that the input is the same as in Fig. \ref{fig:split} and the vector length is 4. First we create an array of $v$ elements (Alg. \ref{alg:pac}, line \ref{alg:pac:init_v}), where each element starts from state $q_0$. The first SIMD $\delta$ operations are performed on the characters at positions 0, 4, 8, and 12, the second SIMD operations are performed on characters at position 1, 5, 9, and 13, and so on. The SIMD loop (Alg. \ref{alg:pac} Line \ref{alg:pac:simd_loop}) that performs the $\delta$ operations is going to be executed $T.length / v + m-1$ times. The estimated speedup according to the vectorization reports (Fig. \ref{fig:vec-report}) is 2.6$\times$ compared to the scalar $\delta$ function.

\subsection {Implementation Aspects}
\label{sec:impl}

\begin{figure}[!ht]
	\centering
	\includegraphics[width=.8\linewidth]{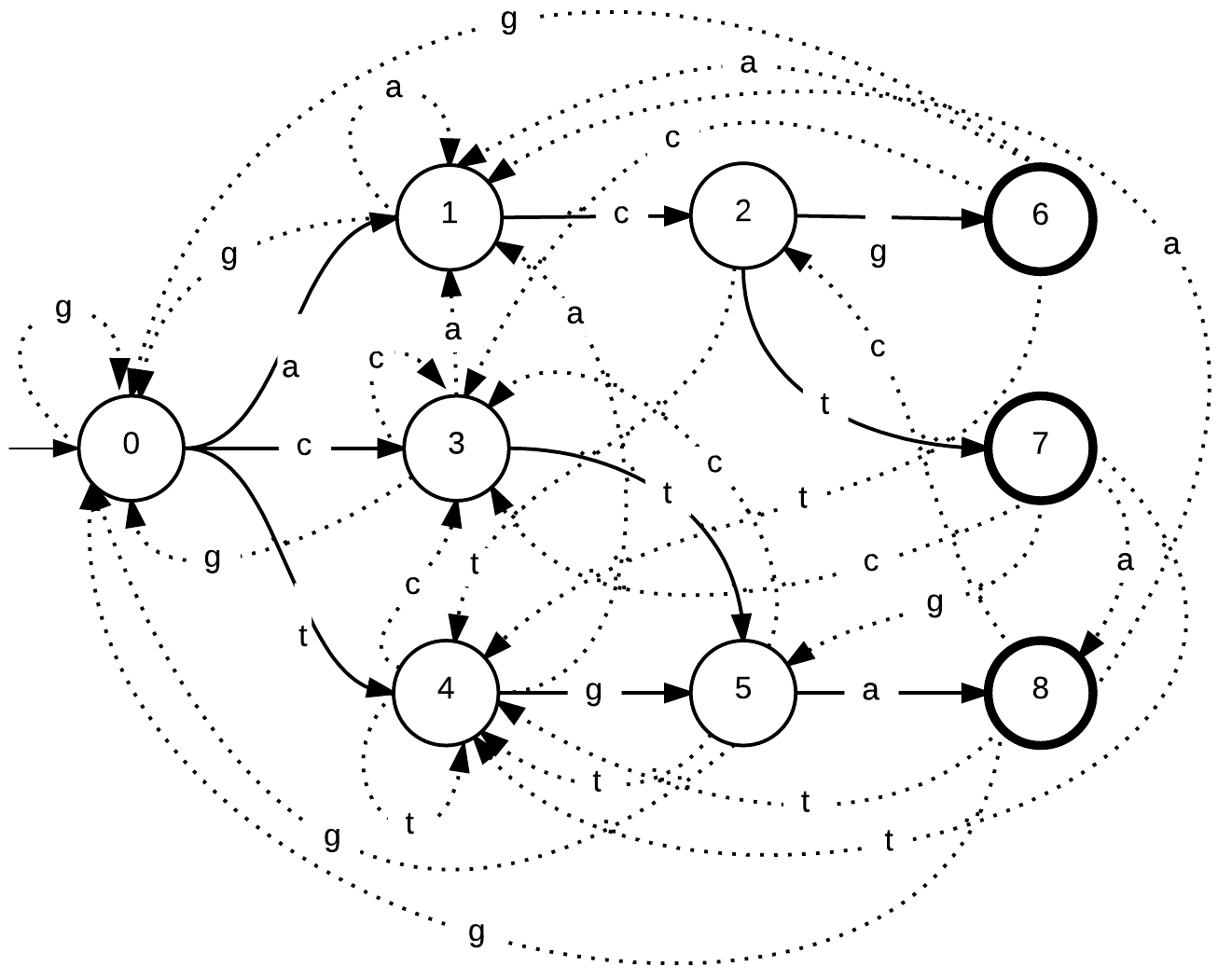}
	\caption{An example of our improved AC automaton that matches occurrences of the following patterns: \emph{acg, act, cta,} and \emph{tga}.}
	\label{fig:ac-dfa}
\end{figure}

The Aho Corasick (AC) with failure links algorithm has a drawback due to its non-deterministic transitions for a single character. Our solution to this issue is illustrated with an example in Figure \ref{fig:ac-dfa}. Our improved AC automaton finds the right transition (indicated by a dashed line) for each state, thus eliminating the failure transitions. Having a valid transition for each possible symbol to another state in the automaton, guarantees that for each character there is always the same number of operations to be performed.

Checking whether the \emph{next state} is a final one requires to store the final states in a set, and then perform a find operation in this set. We have simplified this step by reordering the number of states, such that the regular states are numbered from 0 to $a-b-1$, and the final states are numbered from $a-b$ to $a$, where $a$ is the total number of states and $b$ is the total number of final states. Determining whether a pattern has been found is done by comparing if the \emph{next state} is greater or equal than $a-b$.

A typical representation of the \emph{State Transition Table} (STT) would be a matrix of $x\times y$ elements, where $x$ is the number of states, and $y$ is the size of alphabet $\sum$. The drawback of this representation is that a mapping between the characters of the alphabet and the items on the header of the STT is required (such as, $A\rightarrow0, C\rightarrow1, G\rightarrow2~ and ~T\rightarrow3$). We avoid this issue by representing the automaton as a sparse STT, where the characters of the alphabet are represented by their ASCII code. The size of the STT becomes $x\times z$, where $z$ is the number of items in the ASCII table (that is 256 including the extended ASCII codes). Only the cells that belong to the ASCII codes that represent the characters on the alphabet contain the address to the \emph{next state}, the other ones contain a transition to the start state ($q_0$). 

Table \ref{table:stt} depicts the sparse STT representation of the automaton shown in Fig \ref{fig:ac-dfa}. Our representation of sparse STT is more expensive in terms of memory space, but it is a reasonable trade-off between memory space and access speed.

\begin{table}[!ht]
	\centering
	\caption{STT structure that represents the AC Automaton from Fig. \ref{fig:ac-dfa}}
	\label{table:stt}
	\begin{tabular}{@{}l
			>{\columncolor[HTML]{EFEFEF}}l l
			>{\columncolor[HTML]{EFEFEF}}l l
			>{\columncolor[HTML]{EFEFEF}}l l
			>{\columncolor[HTML]{EFEFEF}}l l
			>{\columncolor[HTML]{EFEFEF}}l @{}}
		\multicolumn{10}{c}{$\leftarrow$ ASCII alphabet (256) $\rightarrow$} \\
		\toprule
		\multicolumn{1}{c}{} & \textit{...} & \textit{A} & \textit{...} & \textit{C} & \textit{...} & \textit{G} & \textit{...} & \textit{T} & \textit{...} \\ \cmidrule(l){2-10} 
		\multicolumn{1}{c}{\multirow{-2}{*}{Q}} & ... & 65 & ... & 67 & ... & 71 & ... & 84 & ... \\ 
		\toprule
		0 &  & $q_1$ &  & $q_3$ &  & $q_0$ &  & $q_4$ &  \\
		1 &  & $q_1$ &  & $q_2$ &  & $q_0$ &  & $q_4$ &  \\
		2 &  & $q_1$ &  & $q_3$ &  & $q_6$ &  & $q_7$ &  \\
		3 &  & $q_1$ &  & $q_3$ &  & $q_0$ &  & $q_5$ &  \\
		4 & ... & $q_1$ & ... & $q_3$ & ... & $q_5$ & ... & $q_4$ & ... \\
		5 &  & $q_8$ &  & $q_3$ &  & $q_0$ &  & $q_4$ &  \\
		6 &  & $q_1$ &  & $q_3$ &  & $q_0$ &  & $q_4$ &  \\
		7 &  & $q_8$ &  & $q_3$ &  & $q_5$ &  & $q_4$ &  \\
		8 &  & $q_1$ &  & $q_2$ &  & $q_0$ &  & $q_4$ &  \\ \bottomrule
	\end{tabular}
\end{table}

\subsection {Algorithm Analysis}

We focus on the most time-consuming parts of our algorithm (Alg. \ref{alg:pac}). The analysis assumes the worst case; for example, the \emph{if-statement} at line \ref{alg:pac:check_final} is assumed to be always true. The estimated time is expressed as follows: 

$T = t_2 + t_3 + t_4 + p*(t_5 + t_6 + t_7) + p*r(t_8 + t_9) + p*r*v(t_10 + t_11 + t_12 + t_13 + t_14)$

where $t_i$ indicates execution time of line $i$, $p$ is the number of processing units, $r$ is the chunk length, and $v$ is the vector length. If $a$ is $t_2 + t_3 + t_4$, $b$ is $t_5 + t_6 + t_7$, $c$ is $t_8 + t_9$, $d$ is $t_10 + t_11 + t_12 + t_13 + t_14$, then we obtain the following,

$T = a + p*b + (p*r)*c + (p*r*v)*d$

Asymptotically we can express the time complexity of our algorithm as:

$\mathcal{O}(p (b + r(c + v * d)))$

The total parallelization overhead of our algorithm can be summarized as: $v(m-1) + p(m-1)$.


\section{Experimental Evaluation} 
\label{exp_evaluation}

In this section we describe the experimentation environment used for the evaluation of our proposed algorithm and we discuss the obtained performance results.

\subsection{Experimentation Environment}

We have implemented our algorithm using C++11 programming language and OpenMP. The algorithm is compiled using the Intel Compiler icc 15.0.0, with enabled O2 optimization option. We have addressed the variability in the performance measurements by repeating the experiment 20 times for each problem size and number of threads. We used the Intel Vtune Amplifier 2015 for performance data collection. 

To evaluate our algorithm the experiments were performed on an Intel Xeon Phi 7120P coprocessor. The Xeon Phi device contains 61 cores, each core supports four hardware threads. The coprocessor software includes the $\mu$OS version 2.6.38.8 and the Intel Manycore Platform Software Stack (MPSS) version 3.1.1. One of the 61 cores is used to run the coprocessor software, and the remaining 60 cores are used for DNA analysis in our experiments. 

For the experimental evaluation we have selected the DNA sequences of mouse, cat, dog, chicken, human and turkey from the GenBank sequence database of the National Center for Biological Information \cite{GenBank}. Information about the genome references and the length of the DNA sequences are listed in Table \ref{table:dna-sequences}.

\begin{table}[h]
		\renewcommand{\arraystretch}{1.3}
		\caption{DNA data-sets}
		\label{table:dna-sequences}
		\centering
		\begin{tabular}{lll}
			\toprule
			& \emph{Gneome Reference}   	 & \emph{Size (MB)} \\ \midrule
			\emph{Mouse}   & GRCm38.p2        		 & 2830      \\ 
			\emph{Cat}     & Felis\_catus-6.2  	 & 2490      \\ 
			\emph{Dog}     & CanFam3.1          	 & 2440      \\ 
			\emph{Chicken} & Gallus\_gullus-4.0	 & 1060      \\ 
			\emph{Human}   & GRCh38            	 & 3250      \\ 
			\emph{Turkey}  & Meleagris\_gallopavo   & 193       \\ \bottomrule
		\end{tabular}
\end{table}
	
In our experiments we use a set of patterns (see Table~\ref{table:regex}) from the \emph{regex-dna} \cite{REGEX_DNA} benchmark that match and extract specific 8-mers from a DNA sequence. 

\begin{table}[ht]
			\renewcommand{\arraystretch}{1.3}
			\caption{Patterns of the \emph{regex-dna} benchmark.}
			\label{table:regex}
			\centering
			\begin{tabular}{ p{2.5cm} p{2.5cm} }
				\hline 
				$agggtaaa$ & $tttaccct$\\
				$(c|g|t)gggtaaa$ & $tttaccc(a|c|g)$\\
				$a(a|c|t)ggtaaa$ & $tttacc(a|g|t)t$\\
				$ag(a|c|t)gtaaa$ & $tttac(a|g|t)ct$\\
				$agg(a|c|t)taaa$ & $ttta(a|g|t)cct$\\
				$aggg(a|c|g)aaa$ & $ttt(c|g|t)ccct$\\
				$agggt(c|g|t)aa$ & $tt(a|c|g)accct$\\
				$agggta(c|g|t)a$ & $t(a|c|g)taccct$\\
				$agggtaa(c|g|t)$ & $(a|c|g)ttaccct$\\
				\hline
			\end{tabular}
\end{table}

\subsection{Results}

We evaluated our algorithm on the Xeon Phi with different numbers of threads for each of the DNA sequences listed in Table \ref{table:dna-sequences}. Furthermore, we have varied the threads affinity, by allocating the threads under \emph{compact, balanced,} and \emph{scatter} mode. The \emph{compact} mode completely fills a core with threads (that is, allocates threads to all available hardware threads of a core) before assigning threads to another core. The \emph{balanced} mode evenly distributes the threads among cores. This mode keeps thread IDs close to each other, which increases the probability that threads on the same core are using data that is close to each other. The \emph{scatter} mode evenly distributes threads among cores in a round-robin fashion, which in contrast to the \emph{balanced} mode assures that the thread IDs are not close to each other \cite{barth2013best}. 

Table \ref{table:affinity} shows the execution time for the three different thread allocation affinity modes. The values in bold indicate the fastest execution time. We can see that for 240 threads, the \emph{balanced} mode is the fastest one for all tested DNA sequences. For 180 threads the \emph{scatter} mode performs the best in the case of dog's DNA sequence.

\begin{figure}
	\centering
	\includegraphics[width= 0.9 \linewidth]{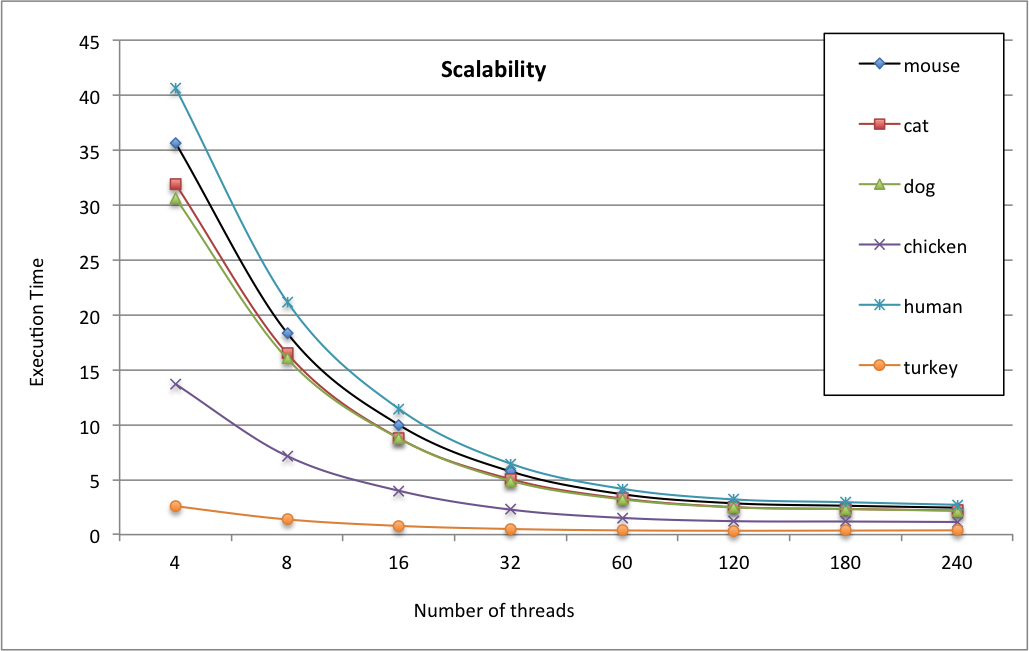}
	\caption{The scalability of our algorithm on the Xeon Phi for various number of threads and problem sizes.}
	\label{fig:execution}

\end{figure}

\begin{figure}
	\centering
	\includegraphics[width= \linewidth]{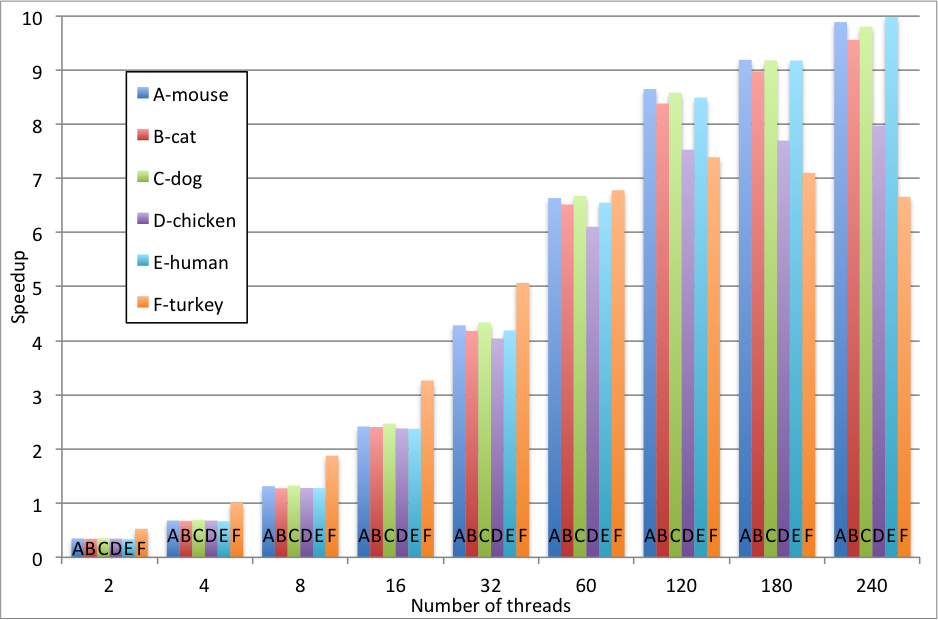}
	\caption{Speedup of our algorithm with respect to a sequential version running on an Intel Xeon E5-2695v2 CPU.}
	\label{fig:speedup}
\end{figure}

The performance data for scalability (Fig. \ref{fig:execution}) and the speedup (Fig. \ref{fig:speedup}) is collected for the \emph{balanced} thread affinity. 

Fig. \ref{fig:execution} shows the scalability of our algorithm when we increase the number of threads on the Intel Xeon Phi coprocessor. Our algorithm scales well up to 120 threads for most of the tested DNA sequences. Increase of the number of threads to 180 or 240, results with a modest performance improvement due to the thread management overhead. The performance gain when using a larger number of threads is higher for larger DNA sequences. Thus the best scalability we observe for the human DNA sequence, which is the largest DNA sequence used in our experiments. 

\begin{table}
	\caption{The execution time on the coprocessor for different number of threads and thread affinity modes.}
	\label{table:affinity}
	\centering
	\begin{tabular}{p{0.65cm}p{0.4cm}p{0.4cm}p{0.4cm}p{0.4cm}p{0.4cm}p{0.4cm}p{0.4cm}p{0.4cm}p{0.4cm}p{0.4cm}}
		\toprule
		Affinity & \multicolumn{3}{c}{Compact} & \multicolumn{3}{c}{Balanced} & \multicolumn{3}{c}{Scatter} \\ \midrule
		Threads & 240 & 180 & 120 & 240 & 180 & 120 & 240 & 180 & 120 \\ \midrule
		Mouse & 2.45 & 2.72 & 3.36 & \textbf{2.44} & \textbf{2.63} & \textbf{2.85} & 2.66 & 2.76 & 3.00 \\
		Cat & 2.21 & 2.43 & 2.99 & \textbf{2.19} & \textbf{2.34} & \textbf{2.51} & 2.27 & 2.35 & 2.52 \\
		Dog & 2.19 & 2.39 & 2.95 & \textbf{2.16} & 2.32 & \textbf{2.48} & 2.27 & \textbf{2.31} & 2.52 \\
		Chicken & 1.18 & 1.23 & 1.43 & \textbf{1.15} & \textbf{1.19} & 1.22 & 1.25 & 1.21 & \textbf{1.21} \\
		Human & 2.77 & 3.07 & 3.82 & \textbf{2.71} & \textbf{2.95} & 3.20 & 2.88 & 2.97 & \textbf{3.19} \\
		Turkey & 0.41 & 0.38 & 0.39 & \textbf{0.39} & \textbf{0.36} & \textbf{0.35} & 0.43 & 0.38 & 0.36 \\ \bottomrule
	\end{tabular}
\end{table}

Fig. \ref{fig:speedup} presents the achieved speedup. The maximal speedup of $10\times$ is achieved for the human DNA sequence using 240 threads compared to a sequential version running on an Intel Xeon E5-2695v2 CPU.

\begin{figure}[!t]
	\centering
	\includegraphics[width= \linewidth]{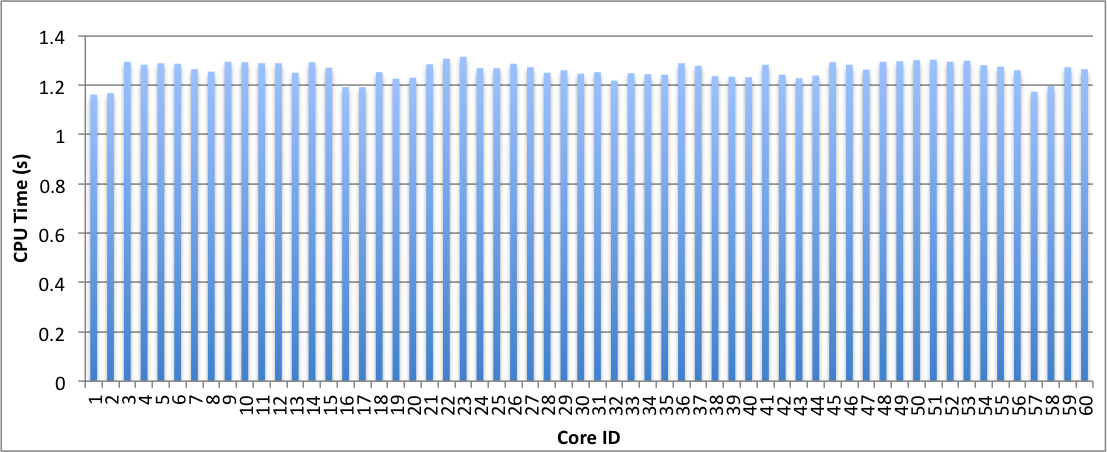}
	\caption{CPU time for each core. The human DNA sequence is processed with the \emph{balanced} thread affinity mode.}
	\label{fig:core-allocation}
\end{figure} 

Figure \ref{fig:core-allocation} shows the CPU time for each core when the human DNA sequence is processed with the \emph{balanced} thread affinity mode. Theoretically we may expect the CPU time of each core to be the same, because each core has the same amount of symbols to process. The CPU time of individual cores may depend not only on the chunk length but also on how many occurrences of the patterns are in the corresponding sequence chunk.


\section{Related State-of-the-Art} 
\label{related_work}

In this section we discuss the state-of-the-art in pattern matching and DNA analysis using many-core architectures (such as, GPU and Intel Xeon Phi).

Villa et al. \cite{VillaCM09} implemented the Aho-Corasick string matching algorithm on a Cray XMT system. Tumeo and Villa implemented the algorithm presented in \cite{VillaCM09} for GPU clusters \cite{Tumeo_DNA_GPU}. Their implementation is based on splitting the input into chunks, and then processing each chunk in a separate thread. In contrast to our approach, their algorithm for pattern matching relies on the features of the Cray XMT or GPU architecture, whereas our algorithm is tailored for DNA analysis on Intel Xeon Phi architecture.

An acceleration of exact string matching Knuth-Morris-Pratt algorithm on GPU is conducted by Bellekens et al. \cite{bellekens2013investigation}. They achieve nearly a $29\times$ speedup compared to the sequential version of the KMP algorithm. Similarly, Kouzinopoulos and Margaritis \cite{kouzinopoulos2009string} conducted an experiment on the Naive, KMP, Boyer-Moore-Horspool and Quick-Search string matching algorithms in the context of DNA sequencing using the CUDA toolkit. In contrast our work addresses large-scale DNA analysis on Intel Xeon Phi.  

Lin et al. \cite{Lin_PFA-C_Algo} evaluated their Parallel Failure-less AC algorithm on GPU and showed improvement of $14.74 \times$. This algorithm allocates a new thread to each character of the input to identify any pattern starting from that character, which means that it creates $n$ number of threads, where $n$ is the input length. In their experiments the length of input string is up to 256MB. While this approach is tailored for pattern matching on GPU, we focus on DNA analysis on Xeon Phi.

Li et al. \cite{wu-manber-cuda} implemented in CUDA the Wu-Manber algorithm, which is used for approximate matching of nucleotides in DNA sequences on GPU. In contrast our algorithm performs exact pattern matching on Intel Xeon Phi.

To the best of our knowledge, our approach for large-scale DNA analysis is the first one that exploits the thread level and SIMD parallelism available on the Intel Xeon Phi coprocessor. In our experiments we have evaluated our approach with real-world DNA sequences of several GB.


\section{Summary and Future Work} 
\label{summary_future_work}

Fast DNA analysis is important in many applications, such as, preventing the evolution of different viruses during an early phase, early diagnosis of genetic predispositions to certain diseases, or DNA forensics. 

In this paper we have presented an approach for accelerating DNA analysis using the Intel Xeon Phi coprocessor. The proposed parallel algorithm is based on finite automata and is used for counting and extracting the location of k-mers in a DNA sequence. Our approach exploits the thread-level and SIMD parallelism of the Intel Xeon Phi coprocessor, and therefore it is suitable for large-scale DNA sequences. Experiments with real-world data-sets of several GB demonstrate that the algorithm scales well with respect to various numbers of threads and problem sizes. The best scalability we observed for the human DNA sequence, which was the largest DNA sequence used in our experiments. 

Future work will address the DNA analysis on the upcoming generation of the Intel Xeon Phi coprocessor known as the \emph{Knights Landing}.

\bibliographystyle{IEEEtran}


\end{document}